      [1999/12/01 v1.4c Il Nuovo Cimento]
\documentclass{cimento}

\usepackage{graphicx}  
\def\kms{km~s$^{-1}$}
\def\etal{{\it et al.}}
\def\sqd{{deg$^{2}$}}

\def\arcsec{$^{\prime\prime}$}
\def\msun{$M_\odot$}
\def\mhi{$M_{HI}$}

\title{ALFALFA: The Search for (Almost) Dark Galaxies and their Space Distribution}
\author{Martha P. Haynes}
\instlist{\inst Center for Radiophysics and Space Research and
       National Astronomy and Ionosphere Center,
       Cornell University, Ithaca
       N.Y. 14853 U.S.A.}
\PACSes{
\PACSit{95.35.+d,}{95.80.+p, }{95.85.Bh, }{98.52.Nr, }{98.52.Sw, }
{98.52.Wz, }{98.58.Ge, }{98.58.Nk, }{98.62.Py, }{98.62.Ve, }{98.80.Es}
}
\begin{document}

\maketitle
\begin{abstract}
The Arecibo Legacy Fast ALFA (ALFALFA) survey is designed to explore the
$z=0$ HI mass function (HIMF) over a cosmologically significant volume. ALFALFA 
will improve on previous determinations of the HIMF by its combination
of depth, wide area and centroiding accuracy, the latter allowing, in most
cases, immediate identification of the optical counterpart to
each HI signal. ALFALFA will detect hundreds of galaxies with HI
masses less than $10^{7.5}$ \msun~ and also greater than $10^{10.5}$ \msun, and
its final catalog will allow investigation of the dependence of 
the HIMF both on local density and on galaxy morphology. Already
ALFALFA confirms previous suggestions that there is no cosmologically
significant population of HI-rich dark galaxies. Fewer than 3\% of all 
extragalactic HI sources and $<$1\% of ones with M$_{HI} > 10^{9.5}$ cannot 
be identified with a stellar counterpart. Very preliminary results on the 
presence of gas-rich dwarfs in the void in front of the Pisces--Perseus 
supercluster suggest an underabundance of such objects compared to the 
predictions of numerical simulations. The objects with highest HI mass 
exhibit a range of morphologies and optical colors and surface brightnesses
but all appear to be massive disk systems. The latter represent the population
likely to dominate future studies of HI at higher redshift with
the Square Kilometer Array.
\end{abstract}

\section{Introduction}
As discussed in the preceding paper by Giovanelli in this volume,
the Arecibo Legacy Fast ALFA (ALFALFA: 
\cite{ref:Giov05a,ref:Giov05b,ref:Giov07,ref:Saint08})
survey is an on-going second
generation HI blind mapping program that will produce a final 
catalog of more than 25000 HI detections at c$z < 0.06$.
The local extragalactic sky visible to Arecibo is rich,
containing the central longitudes of the Supergalactic Plane in and around the
Virgo cluster, the main ridge of the Pisces--Perseus Supercluster,
and the extensive filaments connecting A1367, Coma and Hercules.
With the installation in 2004 of its first ``radio camera'', the 7-beam
Arecibo L-band Feed Array (ALFA),
the Arecibo legacy of extragalactic HI studies continues to
probe regimes untouched by other surveys, addressing
fundamental cosmological questions (the number density, distribution
and nature of low mass halos) and issues of galaxy formation and
evolution (sizes of HI disks, history of tidal interactions and mergers,
low z absorber cross section, origin of dwarf galaxies, nature
of high velocity clouds). In addition to Arecibo's huge collecting area,
the availability of several multi-bit and many-channel backends
allows the {\it simultaneous} collection of multiple datasets
covering different frequency intervals with different spectral
resolution without loss of signal. The combination of Arecibo's
unmatched sensitivity with its new wide area mapping capability
and the efficiency of commensal observing offers tremendous potential 
for future scientific discovery in the coming years.

It is important to note that our current understanding of large
scale structure and the galaxy population traced by HI redshift surveys
is distinctly immature relative to that derived from O/IR redshift surveys.
The previous major blind HI survey, the HI Parkes All-Sky Survey
(HIPASS; \cite{ref:Bar01}), 
surveyed 30000 \sqd~ but
detected only 5000 galaxies with a median redshift of $\sim$2800 \kms. 
As a second generation effort, ALFALFA will improve on these statistics
dramatically. Although it will cover only 7000 \sqd, ALFALFA promises
a final catalog in excess of 25000 HI detections with a median redshift
of $\sim$7800 \kms. Its detection rate is on average 4 to 5
galaxies per square degree; in high density regions, that rate rises to 20
or more. This improvement on HIPASS is driven partly by broader
frequency coverage but mostly by the sheer sensitivity of Arecibo and by
technical advances in signal processing and survey strategy. Furthermore,
the Arecibo beam size advantage permits a centroiding accuracy generally
better than 20\arcsec~ so that optical counterparts can be identified --
or excluded -- with high probability. 
While not designed for the detailed study of the HI distribution within galaxies
and confusion limited in regions of highest local sky density, ALFALFA will
yield a catalog of accurate HI line fluxes, systemic velocities and profile 
widths for gas--rich galaxies within $z < 0.06$, probing a volume
a factor of 10 times greater than that sampled by HIPASS. Initiated in
2005, ALFALFA is expected to be completed in 2010-11. Because the survey is
not complete, results are not yet definitive until more volume is sampled,
but ALFALFA already promises to deliver robust measures of the HI mass function, the HI
correlation function and its bias parameter over a cosmologically significant
volume.  Here, I discuss some of the open issues associated with the number 
density and distribution of gas-rich galaxies and summarize the ALFALFA 
potential to address them.

\section{ALFALFA and the Distribution of Gas-Rich Galaxies}\label{sec:distrib}

It is well known that HI sources are typically associated with star-forming 
disk galaxies, many of which are of low optical surface brightness.  
Spirals in rich clusters, however, exhibit strong HI deficiency, so that in regions of
high X-ray emissivity, gas depletion can reach as high as 90\%.
Thus it may come as no surprise that HI blind surveys may trace the low amplitude
fluctuations in the large scale structure, as cluster populations are simply not 
included among the HI detections. Principal aims of ALFALFA are to explore 
quantitatively how widely dispersed is the HI population in comparison with 
structures traced by O/IR samples and what are the impacts of both morphology and 
density on the HI mass function.

\begin{figure}\label{fig:conem}
\hspace{-2.8cm}
\includegraphics[width = 500pt, height = 500pt]{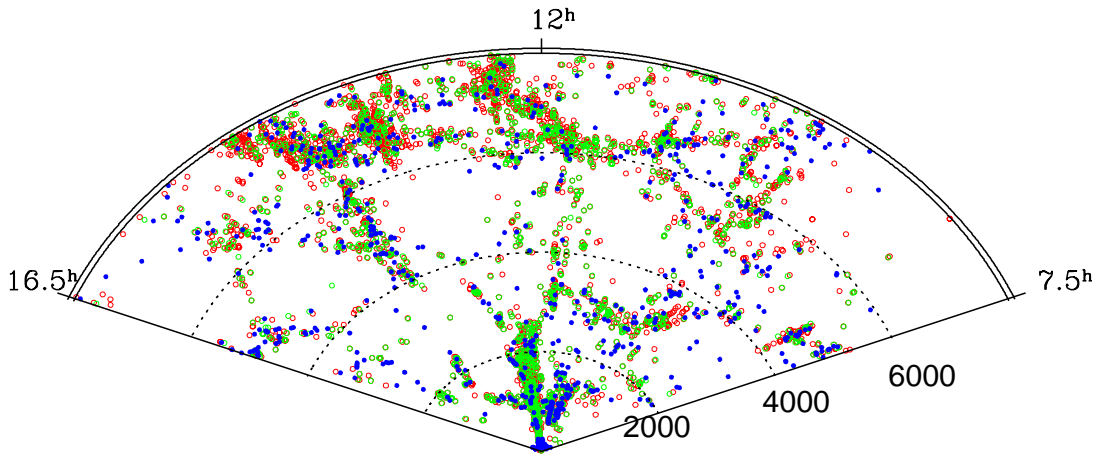}   
\vskip -10.3cm
\caption{Radial distribution of 5670 galaxies with measured
radial velocitites $cz < 8000$ \kms~ in the ALFALFA strip
from $07^h30^m <$ R.A. $< 16^h30^m$ and $+08^{\circ} <$ Decl. $< +16^{\circ}$. Different colors
denote galaxies whose redshifts are drawn from, respectively, optical only (red), 
HI only (blue) and both (green). The HI-rich galaxies trace the same structures seen
by optical surveys; the few ``void'' galaxies seen here are gas rich as expected.
When complete, ALFALFA will provide a statistically
complete picture of the local filament and void population.}
\end{figure}

Figure \ref{fig:conem} shows a cone diagram of the status of redshift
observations in a nearby slice of the sky covered by both ALFALFA and SDSS,
focusing on the local volume out to c$z < 8000$ \kms. The sky area extends
from $07^h30^m <$ R.A. $< 16^h30^m$ and $+08^{\circ} <$ Decl. $< +16^{\circ}$.
Different symbols show the locations of subsets with redshifts derived
from optical observations only (red open circles), HI only (blue filled circles)
and both (green open circles). The central region of this strip is dominated by the
Leo and Virgo regions within the Local Supercluster at c$z < 2000$ \kms,
while the eastern half shows the southern edge of the ``Great Wall'' and
the filaments leading to Coma further to the north. Clearly the optical
surveys, in this case mainly SDSS plus targeted surveys of groups and clusters, dominate the
highest density regions, while the gas-rich galaxies trace the filamentary
structures in more detail. Note that ALFALFA contributes a host of new redshifts
at the nearer distances; these are typically low surface brightness, faint
galaxies which populate the lowest density quartile. 

Both Meyer \etal~\cite{ref:Meyer07} and Basilakos \etal~\cite{ref:Basil07}
have derived the spatial correlation function
$\xi(r)$ for the HIPASS catalog,
employing different methods to account for large scale structure.
On the one hand, Meyer \etal~\cite{ref:Meyer07} conclude that the HI rich population
represented by the HIPASS sample is extremely weakly clustered, though the
clustering scale depends on the galaxy rotational velocity. In contrast,
Basilakos \etal~
\cite{ref:Basil07} claim that the massive HIPASS galaxies show the
same clustering characteristics as optically selected samples, but that
the low mass (M$_{HI} < 10^9$ \msun) systems show a nearly uniform distribution.
A simple explanation for this apparent discrepancy in the analysis of the
same sample is the lack of adequate volume sampling, the effect of which
is to produce different results depending on the nature of corrections
made in accounting for large scale structure.

As Peebles has discussed in this volume, current $\Lambda$CDM simulations predict
that low amplitude filamentary structures criss-cross the voids. The galaxies
corresponding to the halos in those filaments are expected to be low luminosity, 
star forming galaxies \cite{ref:Hoyle05}, excellent candidates 
for detection by ALFALFA. 
A first tantalizing result has been obtained by Saintonge \etal~\cite{ref:Saint08} who
analyzed the ALFALFA catalog covering a portion of the nearby void in 
front of the Pisces-Perseus Supercluster at c$z \sim 2000$ \kms. Within a
volume of 460 Mpc$^{-3}$, ALFALFA detects not a single galaxy. In contrast,
we would have expected to detect 38 HI sources in such a volume based on
scaling the predictions of Gottl\"ober \etal~\cite{ref:Gott03} with a dark--to--HI
mass ratio of 10:1. It is not clear if this discrepancy, based on only
2\% of the ALFALFA catalog, is real or just another example of the perils 
of volume limitations.  Once sufficient volume is sample, ALFALFA
will be able to place stringent constraints on the local void population
and its possible challenge to the $\Lambda$CDM paradigm \cite{ref:Peebles01}.

\section{ALFALFA, the HI Mass Function and ``Missing Satellites''}

Early studies of the HIMF \cite{ref:Briggs93,ref:Hoff92} showed its
functional similarity to the optical luminosity function. Because
of its relevance to the ``missing satellite'' problem, the faint (low
mass) end slope, $\alpha$, of the HIMF has been the focus of many previous studies,
with widely varying results \cite{ref:Zwaan97, ref:RS02, ref:Zwaan03}. 
These previous studies were severely limited in their
sampling of objects at both the low mass ($10^{8}$ \msun) and 
high mass ($10^{10}$ \msun) ends. For example, the Rosenberg \& Schneider\cite{ref:RS02}
sample includes only a dozen galaxies with M$_{HI} < 10^{8}$ 
\msun; the HIPASS result \cite{ref:Zwaan03}
is based on $\sim$ 30 such low mass objects. In addition to the
statistical uncertainties associated with such small samples, the
determination of the faint end slope is complicated by uncertainties
in the distance estimates of these very nearby galaxies.
For example, Masters \etal~\cite{ref:Masters04} 
have shown than the Hubble-law distance model employed by the HIPASS
team \cite{ref:Zwaan03} to their southern hemisphere-only sample yields a 
systematical underestimate of $\alpha$. 

ALFALFA will provide a factor of ten improvement in the volume sampling
over HIPASS, important not only for taking proper account of the impact
of large scale structure but also for separating the impacts of the
morphological variation in HI content \cite{ref:Roberts94} and 
morphological segregation. Discussing particularly the issues associated
with the derivation of the HIMF, Springob \etal~
\cite{ref:Spring05} point out the critical need for
large ``fair'' samples so that corrections can account not just for the
fact that the space density varies with distance but also that the
fractional volume of space occupied by regions of a particular
density do also. Applications of HIMF derivations by methods which are not
sensitive to large scale structure require understanding of the sample
completeness, noise characteristics, RFI impact and signal extraction bias.
Because of its combination of wide areal coverage, depth and angular resolution, 
ALFALFA will sample a sufficiently large volume of the local universe to
yield a cosmologically fair sample.

Although dwarf galaxies constitute the majority of the galaxy population, large 
uncertainties surround their formation and evolutionary histories. 
The dwarf population follows a strong morphology-density relation, with the
passively evolving systems always found within close proximity of massive galaxies,
in contrast to the more widespread gas-rich, star forming population. ALFALFA
will address many questions associated with the widely-dispersed
gas-rich dwarf population: Where are the ``missing satellites''? How are dwarfs 
affected by reionization? Is their evolution dominated by nearby massive neighbors?
Are some dwarfs ``young'', forming only now out of tidal debris?
Environment-dependent mechanisms invoked to drive the morphological segregation include
the tidal and ram pressure stripping as well as the formation of new dwarfs
in the debris of tidal encounters. Indeed, some dwarfs may be recent entities, 
formed out of the tidal debris; these tidal dwarfs are distinctive in having little 
dark matter and higher heavier element abundances, than expected for their luminosities,
an inheritance from their parent galaxies. An example is the ``old'' tidal 
dwarf candidate VCC~2062 \cite{ref:Duc07} included also in the ALFALFA catalog.

Because of its wide areal coverage, sensitivity, and spectral
resolution, ALFALFA is designed to probe the dwarf population over a wide range
of environments, including local voids and the Virgo cluster. 
In addition to providing a robust determination of the low mass HIMF slope,
the detailed study of the morphology, heavy element abundance, stellar
population characteristics, kinematics and space distribution of these
low mass objects will yield clues on their origin and cosmological
importance. Already, ALFALFA has detected more galaxies with $M_{HI} < 10^8$
\msun~ than all other HI surveys combined. 

\section{ALFALFA, Massive Galaxies and Future Prospects with the SKA}

The principal science questions which drive design of the Square Kilometer
Aray require major HI line surveys over a wide range of redshift. In fact,
an advantage that HI line science has over other wavelengths is the potential
to observe HI signals over the redshift window from $z$ = 0 all the
way back to the ``dark ages''. The challenges to undertaking
such studies should not be underestimated, however, both from observing time and 
signal-to-noise considerations as well as from the perspective of an
ever-increasingly polluted radio interference environment. Even allowing for the likely
increase in the gas content with $z$, only the most massive HI galaxies
will be detected in emission at moderate redshift. 

Previous blind HI surveys did not explore sufficient volumes to detect
the highest HI mass galaxies nor to explore time evolution of the HIMF
or the luminosity--rotation velocity scaling relation. Because of the bandwidth
limitations of ALFA and its ``fast'' observing strategy, ALFALFA will not
explore cosmic evolution, but it will establish the $z=0$ abundance of
high mass galaxies. Already, ALFALFA has detected more than twice as many
objects with M$_{HI} > 2.5 \times 10^{10}$ \msun~ than all of the other
previous blind HI surveys combined. Most of these appear to be luminous
disk systems, some with comparable gas and stellar masses. These
systems represent the most gas-rich massive disks and hence the ones most likely
to be detected by current and
near-term programs to detect HI at moderate redshifts. 
In combination with ALFALFA, the GALEX-Arecibo-SDSS
Survey (GASS; D. Schiminovich, P.I.), scheduled to start in 2008, aims to measure the 
fractional gas content in 1000 massive disk galaxies (stellar mass 
M$_* \sim 10^{10}$) in order to explore the physical mechanisms by which
galaxies acquire and retain their gas and convert it into stars and the
influence on those processes of such factors as environment and AGN feedback.

\section{Conclusions}\label{sec:concl}
ALFALFA is an ongoing survey so that its impact is only beginning to
become evident, but it promises to yield $> 25000$ extragalactic
HI detections when it is complete. Its areal detection rate and centroiding
accuracy are significant improvements over earlier surveys, and the vast
majority of HI detections can be identified with optical counterpart with
very high probability. Of the currently available ALFALFA catalog, 
fewer than 3\% of all extragalactic HI sources
and less than 1\% of detections with
\mhi ~$> 10^{9.5}$\msun ~cannot be identified with a stellar component.
When complete, ALFALFA will yield not only valuable measures of gas
content and dynamical mass for several tens of thousands of individual galaxies but also
robust measures of the HIMF, the HI-HI and HI-optical correlation functions 
and their bias parameters at $z = 0$. In combination with surveys at other
wavelengths, the products of ALFALFA will lay a firm footing for 
future studies of the evolution of these parameters over cosmic time.
ALFALFA is an open consortium and interested
parties are invited to follow the survey's progress via the ALFALFA website
{\it http://egg.astro.cornell.edu/alfalfa}.

\acknowledgments
This work has been supported by NSF grants AST--0307661,
AST--0435697 and AST--0607007 and by the Brinson Foundation.
The Arecibo Observatory is part of the National Astronomy and Ionosphere
Center which is operated by Cornell University under a cooperative
agreement with the National Science Foundation. It is a pleasure to
share the excitement of ALFALFA with other members of the survey team
without whose hard work and enthusiastic participation its science could
not be achieved.

\end{document}